\documentclass[aps,twocolumn,showpacs]{revtex4}
\usepackage{epsfig}
\usepackage{longtable}
\newcommand{\nc}{\newcommand}
\nc{\rnc}{\renewcommand}

\nc{\beq}{\begin{equation}}
\nc{\eeq}{\end{equation}}
\nc{\bea}{\begin{eqnarray}}
\nc{\eea}{\end{eqnarray}}
\nc{\ba}{\begin{array}}
\nc{\ea}{\end{array}}
\nc{\bpi}{\begin{picture}}
\nc{\epi}{\end{picture}}
\nc{\nn}{\nonumber}

\nc{\p}{\partial}
\nc{\f}[2]{\frac{#1}{#2}}
\nc{\od}{{\cal O}}
\nc{\ra}{\rightarrow}
\nc{\Rcal}{\cal R}
\nc{\uh}{\hat{u}}

\nc{\be}{\beta}
\nc{\de}{\delta}
\nc{\om}{\omega}
\nc{\ze}{\zeta}
\nc{\De}{\Delta}
\nc{\Si}{\Sigma}

\begin{document}

\bibliographystyle{apsrev}

\title{Bose-Einstein Condensation Temperature of Homogenous Weakly
Interacting Bose Gas in Variational Perturbation Theory Through Six Loops}

\author{Boris Kastening}
%\email[Email address: ]{boris.kastening@physik.fu-berlin.de}
\affiliation{\mbox{Institut f\"ur Theoretische Physik,
Freie Universit\"at Berlin, Arnimallee 14, D-14195 Berlin, Germany}\\
email: {\tt boris.kastening@physik.fu-berlin.de}}

\date{March 2003}

\begin{abstract}
We compute the shift of the transition temperature for a homogenous
weakly interacting Bose gas in leading order in the scattering
length $a$ for given particle density $n$.
Using variational perturbation theory through six loops in a classical
three-dimensional scalar field theory, we obtain
$\De T_c/T_c=1.25\pm0.13\,an^{1/3}$, in agreement with recent
Monte-Carlo results.
\end{abstract}

\maketitle

A dilute homogenous Bose gas with particle density $n$, where the
scattering length $a$ is small compared to the interparticle spacing
$\sim n^{-1/3}$ is a fine example of how, under the right conditions,
collective effects can generate strongly coupled modes from microscopic
degrees of freedom exhibiting non-zero, but arbitrarily weak interactions.
These conditions are met when the temperature is close to the
transition temperature for Bose-Einstein condensation (BEC).
Naive perturbation theory (PT) then breaks down for physical quantities
sensitive to the collective long-wavelength modes.
One of the basic questions is about the nature and size of the shift of
the BEC transition temperature due to the interactions.
Although attempts at the problem have a long history \cite{history,Sto},
only the recent advent of experimental realizations
of BEC in dilute gases have prompted considerable work to finally
solve the problem both qualitatively and quantitatively
\cite{other,GrCeLa,HoGrLa,HoKr,BaBlHoLaVa1,BaBlZi,WiIlKr,ArTo,SCPiRa1,
KaPrSv,ArMo,HoBaBlLa,ArMoTo,BaBlHoLaVa2,SCPiRa2,KnPiRa,BrRa,Kl1}.
Here we treat the case of a homogenous gas as opposed to, {\em e.g.}, the
case of a harmonic trap \cite{GiPiStArTo}.

The appropriate formal framework for the treatment of the many-particle
Schr\"odinger equation describing a gas of identical spin-0 bosons is
non-relativistic $(3+1)$-dimensional field theory.
Being interested only in equilibrium quantities, we switch to imaginary
time $\tau=it$ and consider the Euclidean action
\beq
S_{3+1}
\!=\!\!
\int_0^\be\!\!\!d\tau\!\!\int\!\! d^3x\bigg[\psi^*\!\left(\f{\p}{\p\tau}
{-}\f{1}{2m}\nabla^2{-}\mu\right)\psi
+\f{2\pi a}{m}(\psi^*\psi)^2\bigg],
\eeq
where $m$ is the mass of the bosons and $\mu$ the chemical potential and
we work in units where $k_B=\hbar=1$.
For the description of the physics involving the long-wavelength modes,
the modes with non-zero Matsubara frequencies may be integrated out from
the theory, leaving a three-dimensional classical field theory.
This is, in general, a complicated procedure, but, as a first
approximation, we may simply ignore the dependence of $\psi$ on $\tau$
and obtain the three-dimensional effective Euclidean action
\beq
\label{s3}
S_3=\int d^3x\left[\f{1}{2}|\nabla\phi|^2
+\f{r_{\rm bare}}{2}\phi^2+\f{u}{24}(\phi^2)^2\right]
\eeq
with the identifications $\psi=\sqrt{mT}(\phi_1+i\phi_2)$,
$r_{\rm bare}=-2m\mu$, $u=48\pi amT$.
For details on the matching beyond this approximation and why, for our
purposes, it is indeed justified to use (\ref{s3}), see Ref.\ \cite{ArMoTo}.
It is convenient to generalize the model (\ref{s3}) to an O($N$)
field theory, where $\phi=(\phi_1,\ldots,\phi_N)$,
$\phi^2\equiv\phi_a\phi_a$.
This allows to make contact with the exactly known large-$N$ result
for $\De T_c$ \cite{BaBlZi}.

It has been argued convincingly \cite{BaBlHoLaVa1,BaBlZi,HoBaBlLa}
that the shift of the transition temperature away from the
non-interacting value
\beq
\label{t0n0}
T_0=\f{2\pi}{m}\left[\f{n}{\ze(3/2)}\right]^{2/3}
\eeq
may be expanded as
\beq
\label{deltatexp}
\f{\De T_c}{T_0}=c_1an^{1/3}+[c_2'\ln(an^{1/3})+c_2''](an^{1/3})^2+\cdots
\eeq
and that the constant $c_1$ is independent of the non-zero Matsubara
modes and can be computed within the above three-dimensional field theory
(\ref{s3}) (for the exact value of $c_2'$ and an approximate evaluation
of $c_2''$, see Ref.\ \cite{ArMoTo}).

Apart from the fact that the leading dependence of $\De T_c$ on $a$ has
only been resolved recently \cite{BaBlHoLaVa1}, attempts at the
determination of $c_1$ have produced results from $-0.93$ \cite{WiIlKr}
to $4.7$ \cite{Sto} as has been summarized for instance in Ref.\ \cite{ArMo},
see also Fig.\ \ref{fig2} below.
The various approaches have different weaknesses and we comment on some
of them below.
It appears that the most reliable results so far are obtained by
Monte-Carlo simulations (MC) \cite{ArMo,KaPrSv}.

Improving perturbation theory by a variational principle is an old
idea \cite{Yu}.
Earlier attempts using resummed PT to determine $\De T_c$
\cite{BrRa,SCPiRa1,SCPiRa2,KnPiRa} have employed the linear $\de$
expansion (LDE).
These were criticized in Refs.\ \cite{Kl1,HaKl}, where Kleinert's field
theoretic variational perturbation theory (VPT, see \cite{Kl2,Kl3,Kl4}
and Chapters 5 and 19 of the textbooks \cite{pibook} and \cite{phi4book},
respectively) was used through five loops.
We extend this calculation to six loops with a different treatment
of the one-loop term, which is absent here.
Our first non-zero coefficient arises only at three loops.
In consequence, we obtain a 30\% larger result than Ref.\ \cite{Kl1}.
We comment on the issue at the end of the paragraph following
Eq.\ (\ref{deltaphi2}) below.

After the interaction is turned on, we can either tune the temperature
by $\De T_c$ or the particle density by $\De n$ to bring the system
back to the transition, keeping the respective other quantity fixed.
Eq.\ (\ref{t0n0}) relates the two shifts in leading order
\cite{BaBlHoLaVa1},
\beq
\label{detc}
\f{\De T_c}{T_0}=-\f{2}{3}\f{\De n}{n}.
\eeq
In other words, we may determine $c_1$ by computing $\De n$ at the
critical point.
$\De n$ is related to the interaction-induced shift of
$\langle\phi^2\rangle$ in the three-dimensional theory by
\beq
\label{deltanc}
\De n=\De\langle\psi^*\psi\rangle=mT\De\langle\phi^2\rangle.
\eeq
Combining (\ref{t0n0}), (\ref{deltatexp}), (\ref{detc}) and (\ref{deltanc}),
$c_1$ is given by
\beq
\label{c1}
c_1=-\f{256\pi^3}{[\ze(3/2)]^{4/3}}
\left.\f{\De\langle\phi^2\rangle}{Nu}\right|_{\rm crit.}
\approx-2206.19\left.\f{\De\langle\phi^2\rangle}{Nu}\right|_{\rm crit.}
\eeq
and the remaining task is to compute the critical limit of
$\De\langle\phi^2\rangle/Nu$ in the three-dimensional theory.
We have
\beq
\label{deltaphi2}
\De\langle\phi^2\rangle=N\int\f{d^3p}{(2\pi)^3}[G(p)-G_0(p)],
\eeq
where $G$  and $G_0$ are the interacting and the free propagator,
respectively.
The theory (\ref{s3}) is super-renormalizable and only $r_{\rm bare}$ has
to be tuned to cancel UV divergences (a further divergence in the free
energy is of no interest to us here).
A convenient renormalization scheme defines the renormalized quantity
$r=r_{\rm bare}-\Si(0)$, where $\Si(p)$ is the self-energy.
The full propagator reads then
$G(p)=1/\{p^2+r-[\Si(p)-\Si(0)]\}$
and $r=0$ corresponds to the critical point, where BEC takes place.
Consequently, $G_0(p)=1/(p^2+r)$ \cite{BrRa}.
Another popular scheme is to use $G_0(p)=1/p^2$
\cite{SCPiRa1,SCPiRa2,KnPiRa,Kl1,HaKl}.
This generates an unnatural one-loop contribution to $\De T_c$ even in
the absence of interactions.
Although this contribution vanishes as $r\ra0$, it strongly
influences resummation and is responsible for the small value of $c_1$
in Ref.\ \cite{Kl1}.

Perturbation theory organizes $\De\langle\phi^2\rangle/Nu$ in a loop
expansion
\beq
\label{pertsum}
\f{\De\langle\phi^2\rangle}{Nu}
=\sum_{l=1}^\infty a_lu_r^{l-2},~~~~~~
u_r\equiv\f{Nu}{4\pi r^{1/2}}.
\eeq
At one loop, there is no interaction and therefore the difference in
(\ref{deltaphi2}) vanishes.
There is also no two-loop contribution since our renormalization scheme
subtracts all graphs containing momentum-independent self-energy
insertions and there is no two-loop graph without such an insertion.
This reflects the fact that the one-loop correction of the propagator
merely renormalizes the chemical potential, which does not lead to a shift
in the critical temperature \cite{BaBlHoLaVa1,BaBlHoLaVa2}.
The three-loop contribution is
\beq
\label{a3}
a_3u_r=
\f{1}{Nu}\left[
\rule[-14pt]{0pt}{34pt}
\bpi(30,0)(2,0)
\put(17,3){\circle{24}}
\put(17,3){\oval(24,8)}
\put(5,3){\circle*{2}}
\put(29,3){\circle*{2}}
\put(17,-9){\makebox(0,0){$\times$}}
\epi
-
\rule[-14pt]{0pt}{34pt}
\bpi(34,0)
\put(17,3){\circle{24}}
\put(17,3){\oval(24,8)}
\put(5,3){\circle*{2}}
\put(29,3){\circle*{2}}
\put(17,-9){\makebox(0,0){$\times$}}
\put(2,-3){\dashbox{2}(30,20){}}
\epi
\right]
=-\f{(1+\f{2}{N})\ln\f{4}{3}}{576\pi^2}u_r,
\eeq
where the dashed box means that the enclosed self-energy contribution
has to be taken at zero external momentum.
The three-loop case exhibits the only type of divergence appearing
at any loop order of the calculation, the ``sunset'' subdivergence
which causes the two diagrams in (\ref{a3}) to be UV divergent.
One may either regulate each diagram and remove the regulator from the
finite difference or, much simpler, perform the momentum integrations
only after the subtraction to arrive at the given result.

To save space, it is convenient to define an operator $\Rcal$ that
recursively subtracts the zero-momentum contributions of all self-energy
insertions of a given diagram.
E.g., the combination of diagrams in Eq.\ (\ref{a3}) is then represented
by $\Rcal$ applied to the first diagram. 
The diagrams through six loops, their weights and O($N$) symmetry factors
as well as the numerical values of the corresponding integrals are given
in Table \ref{diagrams}.
\setlength{\unitlength}{0.8pt}
\begin{table*}
\begin{center}
\begin{tabular}{|c|c|c|}\hline
\multicolumn{1}{|l|}{$L\mbox{-}n$} &
$w_{L\mbox{-}n}$ & $g_{L\mbox{-}n}$ \\\cline{2-3}
diagram & \multicolumn{2}{c|}{$I_{L\mbox{-}n}$}
\\\hline\hline
\multicolumn{1}{|l|}{\rule[-5pt]{0pt}{15pt}\bpi(0,0)\put(0,3){3-1}\epi}
& $\f{1}{6}$ & $\f{N(N+2)}{3}$
\\\cline{2-3}\rule[-4pt]{0pt}{12pt}
$
%\rule[-14pt]{0pt}{34pt}
%\bpi(44,0)(-10,8)
\bpi(40,0)(-10,-10)
\put(0,3){\makebox(0,0)[r]{$\Rcal$}}
\put(17,3){\circle{24}}
\put(17,3){\oval(24,8)}
\put(5,3){\circle*{2}}
\put(29,3){\circle*{2}}
\put(17,-9){\makebox(0,0){$\times$}}
\epi$
& \multicolumn{2}{c|}{$-7.24858\times10^{-5}$}
\\\hline
\multicolumn{1}{|l|}{\rule[-5pt]{0pt}{15pt}\bpi(0,0)\put(0,3){4-1}\epi}
& $\f{1}{4}$ & $\f{N(N+2)(N+8)}{27}$
\\\cline{2-3}\rule[-4pt]{0pt}{12pt}
$
%\rule[-14pt]{0pt}{34pt}
%\bpi(44,0)(-10,8)
\bpi(40,0)(-10,-10)
\put(0,3){\makebox(0,0)[r]{$\Rcal$}}
\put(17,3){\circle{24}}
\put(6.6,-3){\line(1,0){20.8}}
\put(6.6,-3){\line(3,5){10.4}}
\put(27.4,-3){\line(-3,5){10.4}}
\put(6.6,-3){\circle*{2}}
\put(27.4,-3){\circle*{2}}
\put(17,15){\circle*{2}}
\put(17,-9){\makebox(0,0){$\times$}}
\epi$
& \multicolumn{2}{|c|}{$-2.04919\times10^{-6}$}
\\\hline
\multicolumn{1}{|l|}{\rule[-5pt]{0pt}{17pt}\bpi(0,0)\put(0,5){5-1}\epi}
& $\f{1}{8}$ & $\f{N(N+2)(N^2+6N+20)}{81}$
\\\cline{2-3}\rule[-4pt]{0pt}{12pt}
$
%\rule[-14pt]{0pt}{34pt}
%\bpi(44,0)(-10,8)
\bpi(40,0)(-10,-10)
\put(0,3){\makebox(0,0)[r]{$\Rcal$}}
\put(17,3){\circle{24}}
\put(8.5,-5.5){\line(1,0){17}}
\put(8.5,-5.5){\line(0,1){17}}
\put(8.5,11.5){\line(1,0){17}}
\put(25.5,-5.5){\line(0,1){17}}
\put(8.5,-5.5){\circle*{2}}
\put(8.5,11.5){\circle*{2}}
\put(25.5,-5.5){\circle*{2}}
\put(25.5,11.5){\circle*{2}}
\put(17,-9){\makebox(0,0){$\times$}}
\epi$
& \multicolumn{2}{c|}{$-6.32139\times10^{-8}$}
\\\hline
\multicolumn{1}{|l|}{\rule[-5pt]{0pt}{17pt}\bpi(0,0)\put(0,5){5-2}\epi}
& $\f{1}{12}$ & $\f{N(N+2)^2}{9}$
\\\cline{2-3}\rule[-6pt]{0pt}{16pt}
$
%\rule[-18pt]{0pt}{42pt}
%\bpi(60,0)(-10,10)
\bpi(56,0)(-10,-10)
\put(0,3){\makebox(0,0)[r]{$\Rcal$}}
\put(13,3){\circle{16}}
\put(37,3){\circle{16}}
\put(25,11){\oval(24,16)[t]}
\put(25,-5){\oval(24,16)[b]}
\put(13,-5){\line(0,1){16}}
\put(37,-5){\line(0,1){16}}
\put(13,-5){\circle*{2}}
\put(13,11){\circle*{2}}
\put(37,-5){\circle*{2}}
\put(37,11){\circle*{2}}
\put(5,3){\makebox(0,0){$\times$}}
\epi
$
& \multicolumn{2}{c|}{$4.20763\times10^{-9}$}
\\\hline
\multicolumn{1}{|l|}{\rule[-5pt]{0pt}{17pt}\bpi(0,0)\put(0,5){5-3}\epi}
& $\f{1}{36}$ & $\f{N(N+2)^2}{9}$
\\\cline{2-3}\rule[-6pt]{0pt}{16pt}
$
%\rule[-18pt]{0pt}{42pt}
%\bpi(60,0)(-10,10)
\bpi(56,0)(-10,-10)
\put(0,3){\makebox(0,0)[r]{$\Rcal$}}
\put(13,3){\circle{16}}
\put(37,3){\circle{16}}
\put(25,11){\oval(24,16)[t]}
\put(25,-5){\oval(24,16)[b]}
\put(13,-5){\line(0,1){16}}
\put(37,-5){\line(0,1){16}}
\put(13,-5){\circle*{2}}
\put(13,11){\circle*{2}}
\put(37,-5){\circle*{2}}
\put(37,11){\circle*{2}}
\put(25,-13){\makebox(0,0){$\times$}}
\epi$
& \multicolumn{2}{c|}{\rule[-5pt]{0pt}{15pt}$2.32536\times10^{-9}$}
\\\hline
\end{tabular}
\nolinebreak
\begin{tabular}{|c|c|c|}\hline
\multicolumn{1}{|l|}{\rule[-5pt]{0pt}{17pt}\bpi(0,0)\put(0,5){5-4}\epi}
& $\f{1}{4}$ & $\f{N(N+2)(5N+22)}{81}$
\\\cline{2-3}\rule[-3pt]{0pt}{13pt}
$
%\bpi(56,0)(-10,-10)
\bpi(52,0)(-10,-12)
\put(0,3){\makebox(0,0)[r]{$\Rcal$}}
\put(13,3){\circle{16}}
\put(37,3){\circle{16}}
\put(45,-5){\line(0,1){16}}
\put(29,-5){\oval(32,16)[b]}
\put(29,11){\oval(32,16)[t]}
\put(13,3){\oval(32,16)[r]}
\put(45,3){\circle*{2}}
\put(29,3){\circle*{2}}
\put(13,-5){\circle*{2}}
\put(13,11){\circle*{2}}
\put(5,3){\makebox(0,0){$\times$}}
\epi$
& \multicolumn{2}{c|}{$-3.47299\times10^{-8}$}
\\\hline
\multicolumn{1}{|l|}{\rule[-5pt]{0pt}{17pt}\bpi(0,0)\put(0,5){5-5}\epi}
& $\f{1}{4}$ & $\f{N(N+2)(5N+22)}{81}$
\\\cline{2-3}\rule[-3pt]{0pt}{13pt}
$
%\bpi(56,0)(-10,-10)
\bpi(52,0)(-10,-12)
\put(0,3){\makebox(0,0)[r]{$\Rcal$}}
\put(13,3){\circle{16}}
\put(37,3){\circle{16}}
\put(45,-5){\line(0,1){16}}
\put(29,-5){\oval(32,16)[b]}
\put(29,11){\oval(32,16)[t]}
\put(13,3){\oval(32,16)[r]}
\put(45,3){\circle*{2}}
\put(29,3){\circle*{2}}
\put(13,-5){\circle*{2}}
\put(13,11){\circle*{2}}
\put(29,-13){\makebox(0,0){$\times$}}
\epi$
& \multicolumn{2}{c|}{$-5.39050\times10^{-8}$}
\\\hline
\multicolumn{1}{|l|}{\rule[-5pt]{0pt}{17pt}\bpi(0,0)\put(0,5){6-1}\epi}
& $\f{1}{16}$ & $\f{N(N+2)(N^3+8N^2+24N+48)}{243}$
\\\cline{2-3}\rule[-6pt]{0pt}{16pt}
$
%\rule[-20pt]{0pt}{46pt}
%\bpi(56,0)(-10,10)
\bpi(52,0)(-10,-10)
\put(0,3){\makebox(0,0)[r]{$\Rcal$}}
\put(23,3){\circle{36}}
\put(33.58,-11.56){\circle*{2}}
\put(40.12,8.56){\circle*{2}}
\put(23,21){\circle*{2}}
\put(5.88,8.56){\circle*{2}}
\put(12.41,-11.56){\circle*{2}}
\put(12.41,-11.56){\line(1,0){21.17}}
\qbezier(33.58,-11.56)(36.85,-1.5)(40.12,8.56)
\qbezier(40.12,8.56)(31.56,14.78)(23,21)
\qbezier(23,21)(14.44,14.78)(5.88,8.56)
\qbezier(5.88,8.56)(9.15,1.5)(12.41,-11.56)
\put(23,-15){\makebox(0,0){$\times$}}
\epi$
& \multicolumn{2}{c|}{$-2.04830\times10^{-9}$}
\\\hline
\multicolumn{1}{|l|}{\rule[-5pt]{0pt}{17pt}\bpi(0,0)\put(0,5){6-2}\epi}
& $\f{1}{8}$ & $\f{N(N+2)^2(N+8)}{81}$
\\\cline{2-3}\rule[-5pt]{0pt}{14pt}
$
%\rule[-18pt]{0pt}{42pt}
%\bpi(66,0)(-10,7)
\bpi(62,0)(-10,-10)
\put(0,3){\makebox(0,0)[r]{$\Rcal$}}
\put(13,3){\circle{16}}
\put(35,-5){\circle{16}}
\put(35,11){\circle{16}}
\put(35,-5){\oval(44,16)[bl]}
\put(35,11){\oval(44,16)[tl]}
\put(13,-5){\line(0,1){16}}
\put(35,3){\oval(32,32)[r]}
\put(13,-5){\circle*{2}}
\put(13,11){\circle*{2}}
\put(35,-13){\circle*{2}}
\put(35,3){\circle*{2}}
\put(35,19){\circle*{2}}
\put(5,3){\makebox(0,0){$\times$}}
\epi
$
& \multicolumn{2}{c|}{$1.83919\times10^{-10}$}
\\\hline
\multicolumn{1}{|l|}{\rule[-5pt]{0pt}{17pt}\bpi(0,0)\put(0,5){6-3}\epi}
& $\f{1}{12}$ & $\f{N(N+2)^2(N+8)}{81}$
\\\cline{2-3}\rule[-4pt]{0pt}{14pt}
$
%\rule[-18pt]{0pt}{42pt}
%\bpi(66,0)(-10,7)
\bpi(62,0)(-10,-12)
\put(0,3){\makebox(0,0)[r]{$\Rcal$}}
\put(13,3){\circle{16}}
\put(35,-5){\circle{16}}
\put(35,11){\circle{16}}
\put(35,-5){\oval(44,16)[bl]}
\put(35,11){\oval(44,16)[tl]}
\put(13,-5){\line(0,1){16}}
\put(35,3){\oval(32,32)[r]}
\put(13,-5){\circle*{2}}
\put(13,11){\circle*{2}}
\put(35,-13){\circle*{2}}
\put(35,3){\circle*{2}}
\put(35,19){\circle*{2}}
\put(22,-13){\makebox(0,0){$\times$}}
\epi$
& \multicolumn{2}{c|}{$8.98169\times10^{-11}$}
\\\hline
\multicolumn{1}{|l|}{\rule[-5pt]{0pt}{17pt}\bpi(0,0)\put(0,5){6-4}\epi}
& $\f{1}{24}$ & $\f{N(N+2)^2(N+8)}{81}$
\\\cline{2-3}\rule[-3pt]{0pt}{13pt}
$
%\rule[-18pt]{0pt}{42pt}
%\bpi(66,0)(-10,7)
\bpi(62,0)(-10,-12)
\put(0,3){\makebox(0,0)[r]{$\Rcal$}}
\put(13,3){\circle{16}}
\put(35,-5){\circle{16}}
\put(35,11){\circle{16}}
\put(35,-5){\oval(44,16)[bl]}
\put(35,11){\oval(44,16)[tl]}
\put(13,-5){\line(0,1){16}}
\put(35,3){\oval(32,32)[r]}
\put(13,-5){\circle*{2}}
\put(13,11){\circle*{2}}
\put(35,-13){\circle*{2}}
\put(35,3){\circle*{2}}
\put(35,19){\circle*{2}}
\put(51,3){\makebox(0,0){$\times$}}
\epi$
& \multicolumn{2}{c|}{$1.40475\times10^{-10}$}
\\\hline
\multicolumn{1}{|l|}{\rule[-5pt]{0pt}{17pt}\bpi(0,0)\put(0,5){6-5}\epi}
& $\f{1}{6}$ & $\f{N(N+2)^2(N+8)}{81}$
\\\cline{2-3}\rule[-3pt]{0pt}{13pt}
$
%\rule[-18pt]{0pt}{42pt}
%\bpi(66,0)(-10,7)
\bpi(62,0)(-10,-12)
\put(0,3){\makebox(0,0)[r]{$\Rcal$}}
\put(13,3){\circle{16}}
\put(35,-5){\circle{16}}
\put(35,11){\circle{16}}
\put(35,-5){\oval(44,16)[bl]}
\put(35,11){\oval(44,16)[tl]}
\put(13,-5){\line(0,1){16}}
\put(35,3){\oval(32,32)[r]}
\put(13,-5){\circle*{2}}
\put(13,11){\circle*{2}}
\put(35,-13){\circle*{2}}
\put(35,3){\circle*{2}}
\put(35,19){\circle*{2}}
\put(27,-5){\makebox(0,0){$\times$}}
\epi$
& \multicolumn{2}{c|}{$1.23133\times10^{-10}$}
\\\hline
\end{tabular}
\nolinebreak
\begin{tabular}{|c|c|c|}\hline
%$L\mbox{-}n$ & diagram & 
%$w_{L\mbox{-}n}$ & $g_{L\mbox{-}n}$ \\\cline{2-3}
%&& \multicolumn{2}{c|}{$I_{L\mbox{-}n}$}
%\\\hline\hline\rule[-5pt]{0pt}{15pt}
\multicolumn{1}{|l|}{\rule[-5pt]{0pt}{17pt}\bpi(0,0)\put(0,5){6-6}\epi}
& $\f{1}{4}$ & $\f{N(N+2)(3N^2+22N+56)}{243}$
\\\cline{2-3}\rule[-4pt]{0pt}{14pt}
$
%\rule[-18pt]{0pt}{42pt}
%\bpi(84,0)(-10,8)
\bpi(72,0)(-10,-12)
\put(0,3){\makebox(0,0)[r]{$\Rcal$}}
\put(53,3){\circle{16}}
\put(37,3){\circle{16}}
\put(13,3){\circle{16}}
\put(61,-5){\line(0,1){16}}
\put(37,-5){\oval(48,16)[b]}
\put(37,11){\oval(48,16)[t]}
\put(13,3){\oval(32,16)[r]}
\put(61,3){\circle*{2}}
\put(45,3){\circle*{2}}
\put(29,3){\circle*{2}}
\put(13,-5){\circle*{2}}
\put(13,11){\circle*{2}}
\put(37,-5){\makebox(0,0){$\times$}}
\epi$
& \multicolumn{2}{c|}{$-1.10895\times10^{-9}$}
\\\hline
\multicolumn{1}{|l|}{\rule[-5pt]{0pt}{17pt}\bpi(0,0)\put(0,5){6-7}\epi}
& $\f{1}{4}$ & $\f{N(N+2)(3N^2+22N+56)}{243}$
\\\cline{2-3}\rule[-4pt]{0pt}{14pt}
$
\bpi(72,0)(-10,-12)
\put(0,3){\makebox(0,0)[r]{$\Rcal$}}
\put(53,3){\circle{16}}
\put(37,3){\circle{16}}
\put(13,3){\circle{16}}
\put(61,-5){\line(0,1){16}}
\put(37,-5){\oval(48,16)[b]}
\put(37,11){\oval(48,16)[t]}
\put(13,3){\oval(32,16)[r]}
\put(61,3){\circle*{2}}
\put(45,3){\circle*{2}}
\put(29,3){\circle*{2}}
\put(13,-5){\circle*{2}}
\put(13,11){\circle*{2}}
\put(37,-13){\makebox(0,0){$\times$}}
\epi$
& \multicolumn{2}{c|}{$-1.65022\times10^{-9}$}
\\\hline
\multicolumn{1}{|l|}{\rule[-5pt]{0pt}{17pt}\bpi(0,0)\put(0,5){6-8}\epi}
& $\f{1}{8}$ & $\f{N(N+2)(3N^2+22N+56)}{243}$
\\\cline{2-3}\rule[-4pt]{0pt}{14pt}
$
\bpi(72,0)(-10,-12)
\put(0,3){\makebox(0,0)[r]{$\Rcal$}}
\put(53,3){\circle{16}}
\put(37,3){\circle{16}}
\put(13,3){\circle{16}}
\put(61,-5){\line(0,1){16}}
\put(37,-5){\oval(48,16)[b]}
\put(37,11){\oval(48,16)[t]}
\put(13,3){\oval(32,16)[r]}
\put(61,3){\circle*{2}}
\put(45,3){\circle*{2}}
\put(29,3){\circle*{2}}
\put(13,-5){\circle*{2}}
\put(13,11){\circle*{2}}
\put(5,3){\makebox(0,0){$\times$}}
\epi$
& \multicolumn{2}{c|}{$-8.05795\times10^{-10}$}
\\\hline
\multicolumn{1}{|l|}{\rule[-5pt]{0pt}{17pt}\bpi(0,0)\put(0,5){6-9}\epi}
& $\f{1}{2}$ & $\f{N(N+2)(N^2+20N+60)}{243}$
\\\cline{2-3}\rule[-3pt]{0pt}{12pt}
$
%\rule[-17pt]{0pt}{40pt}
%\bpi(50,0)(-10,7)
\bpi(46,0)(-10,-12)
\put(0,3){\makebox(0,0)[r]{$\Rcal$}}
\put(20,3){\circle{30}}
\put(20,3){\circle*{2}}
\put( 9.39,-7.61){\circle*{2}}
\put( 9.39,13.61){\circle*{2}}
\put(30.61,-7.61){\circle*{2}}
\put(30.61,13.61){\circle*{2}}
\put( 9.39,-7.61){\line(1, 1){21.22}}
\put( 9.39,13.61){\line(1,-1){21.22}}
\put( 9.39,-7.61){\line(1, 0){21.22}}
\put( 9.39,13.61){\line(1, 0){21.22}}
\put(21,-12){\makebox(0,0){$\times$}}
\epi$
& \multicolumn{2}{c|}{$-6.27851\times10^{-10}$}
\\\hline
\multicolumn{1}{|l|}{\rule[-5pt]{0pt}{17pt}\bpi(0,0)\put(0,5){6-10}\epi}
& $\f{1}{2}$ & $\f{N(N+2)(N^2+20N+60)}{243}$
\\\cline{2-3}\rule[-3pt]{0pt}{12pt}
$
%\rule[-17pt]{0pt}{40pt}
%\bpi(50,0)(-10,7)
\bpi(46,0)(-10,-12)
\put(0,3){\makebox(0,0)[r]{$\Rcal$}}
\put(20,3){\circle{30}}
\put(20,3){\circle*{2}}
\put( 9.39,-7.61){\circle*{2}}
\put( 9.39,13.61){\circle*{2}}
\put(30.61,-7.61){\circle*{2}}
\put(30.61,13.61){\circle*{2}}
\put( 9.39,-7.61){\line(1, 1){21.22}}
\put( 9.39,13.61){\line(1,-1){21.22}}
\put( 9.39,-7.61){\line(1, 0){21.22}}
\put( 9.39,13.61){\line(1, 0){21.22}}
\put(14.70,4.30){\line(0,1){8}}
\put(10.70,8.30){\line(1,0){8}}
\epi$
& \multicolumn{2}{c|}{$-1.06164\times10^{-9}$}
\\\hline
\multicolumn{1}{|l|}{\rule[-5pt]{0pt}{17pt}\bpi(0,0)\put(0,5){6-11}\epi}
& $\f{1}{4}$ & $\f{N(N+2)(N^2+20N+60)}{243}$
\\\cline{2-3}\rule[-3pt]{0pt}{12pt}
$
%\rule[-17pt]{0pt}{40pt}
%\bpi(50,0)(-10,7)
\bpi(46,0)(-10,-12)
\put(0,3){\makebox(0,0)[r]{$\Rcal$}}
\put(20,3){\circle{30}}
\put(20,3){\circle*{2}}
\put( 9.39,-7.61){\circle*{2}}
\put( 9.39,13.61){\circle*{2}}
\put(30.61,-7.61){\circle*{2}}
\put(30.61,13.61){\circle*{2}}
\put( 9.39,-7.61){\line(1, 1){21.22}}
\put( 9.39,13.61){\line(1,-1){21.22}}
\put( 9.39,-7.61){\line(1, 0){21.22}}
\put( 9.39,13.61){\line(1, 0){21.22}}
\put(5,3){\makebox(0,0){$\times$}}
\epi$
& \multicolumn{2}{c|}{$-1.32467\times10^{-9}$}
\\\hline
\multicolumn{1}{|l|}{\rule[-5pt]{0pt}{17pt}\bpi(0,0)\put(0,5){6-12}\epi}
& $\f{1}{6}$ & $\f{N(N+2)(5N+22)}{81}$
\\\cline{2-3}\rule[-5pt]{0pt}{17pt}
$
%\rule[-20pt]{0pt}{46pt}
%\bpi(56,0)(-10,9)
\bpi(52,0)(-10,-12)
\put(0,3){\makebox(0,0)[r]{$\Rcal$}}
\put(23,3){\circle{36}}
\put(33.58,-11.56){\circle*{2}}
\put(40.12,8.56){\circle*{2}}
\put(23,21){\circle*{2}}
\put(5.88,8.56){\circle*{2}}
\put(12.41,-11.56){\circle*{2}}
\put(5.88,8.56){\line(1,0){34.24}}
\qbezier(33.58,-11.56)(19.73,-1.5)(5.88,8.56)
\qbezier(33.58,-11.56)(28.29, 4.72)(23,21)
\qbezier(12.41,-11.56)(26.27,-1.5)(40.12,8.56)
\qbezier(12.41,-11.56)(17.71,4.72)(23,21)
\put(23,-15){\makebox(0,0){$\times$}}
\epi$
& \multicolumn{2}{c|}{$-4.25366\times10^{-10}$}
\\\hline
\end{tabular}
\caption{\label{diagrams}
Diagrams from three through six loops, their weights  $w_{L\mbox{-}n}$,
$O(N)$ symmetry factors $g_{L\mbox{-}n}$ and numerical results
$I_{L\mbox{-}n}$ for the corresponding integrals for $r=1$.
The contribution of each diagram to $\langle\phi^2\rangle$
is $(-u)^{L-1}r^{1-L/2}w_{L\mbox{-}n}g_{L\mbox{-}n}I_{L\mbox{-}n}$.}
\end{center}
\end{table*}
The latter were obtained by own computations and confirmed by extracting
them from Ref.\ \cite{NiMeBaMuNi}.
Details will be presented elsewhere \cite{Ka}.
The values for the diagrams through five loops are consistent with those
in Ref.\ \cite{BrRa} [we have more precise results for diagrams 5-4 and
5-5, though;
the five-loop results of \cite{SCPiRa2} have to be taken with care
since at least two of the diagrams (5-2 and 5-3) are wrong there, while
the corresponding coefficient $a_5$ is off by 10\%].

The resulting perturbative coefficients are
$a_1=a_2=0$,
$a_3=-5.06047\times10^{-5}$,
$a_4= 2.99626\times10^{-6}$,
$a_5=-1.93583\times10^{-7}$,
$a_6= 1.31373\times10^{-8}$
for $N\ra\infty$ and
$a_1=a_2=0$,
$a_3=-1.01209\times10^{-4}$,
$a_4= 2.99626\times10^{-5}$,
$a_5=-1.19872\times10^{-5}$,
$a_6= 5.85519\times10^{-6}$
for $N=2$.

To extract the $r\ra0$ limit of (\ref{pertsum}), we need to
apply a resummation algorithm.
The alternating signs of the coefficients $a_l$ suggest to use
a Borel-type method, but since we have no information
on the distance of the first singularity of the Borel sum from the
origin in the complex plane, we proceed differently and use VPT.
This method has proven to yield accurate results for critical exponents
\cite{Kl3,Kl4,phi4book} and amplitude ratios \cite{KlvdB} in the present
context of critical phenomena.
The crucial attractive feature of VPT for our purpose is that it allows
to implement the correct Wegner exponent of approach to scaling
$\om$ [$=\be'(g^*)$ in a renormalization group treatment] \cite{We}.
For quantities like $\De\langle\phi^2\rangle/Nu$ that remain
finite at the critical point, this approach has, in our renormalization
scheme, an expansion
\beq
\f{\De\langle\phi^2\rangle}{Nu}=\sum_{m=0}^\infty f_mu_r^{-m\om'},
\eeq
where $\om'=2\om/(2-\eta)$ and $\eta$ is the anomalous dimension of the
critical propagator.

For a truncated partial sum $\sum_{l=1}^La_lu_r^{l-2}$ of (\ref{pertsum}),
\mbox{Kleinert's} VPT requires replacing
\bea
u_r^{l-2}
\!&\ra&\!
(t\uh)^{l-2}
\left\{1+t\left[\left(\f{\uh}{u_r}\right)^{\om'}-1\right]
\right\}^{-(l-2)/\om'}
\eea
(note that this is an identity for $t=1$),
reexpanding the resulting expression in $t$ through $t^{L-2}$, setting
$t=1$ and then optimizing in $\uh$, where optimizing is done in accordance
with the principle of minimal sensitivity \cite{Ste} and in practice means
finding appropriate stationary or turning points.
That is, we replace
\beq
\label{uuhat}
u_r^{l-2}
\ra
\uh^{l-2}\sum_{k=0}^{L-l}
\left(\ba{c}-(l-2)/\om'\\k\ea\right)
\left[\left(\f{\uh}{u_r}\right)^{\om'}-1\right]^k
\eeq
and optimize the resulting expression in $\uh$.
For $u_r\ra\infty$, we obtain the $L$-loop approximation
of the $r\ra0$ limit of $\De\langle\phi^2\rangle/Nu$,
\beq
\label{f0}
f_0^{(L)}={\rm opt}_{\uh}\left[\sum_{l=1}^La_l\uh^{l-2}\sum_{k=0}^{L-l}
\left(\ba{c}-(l-2)/\om'\\k\ea\right)(-1)^k\right].
\eeq
Results are only available starting at four loops, since two non-zero
perturbative coefficients are necessary for VPT to work.

\begin{figure}[ht]
\begin{center}
% becL40largeNplot.eps produced by running ~/bec/math/resum under Mathematica
%\bpi(10,0)
%\put(0,100){$\De\langle\phi^2\rangle/Nu$}
%\epi
\includegraphics[width=8cm,angle=0]{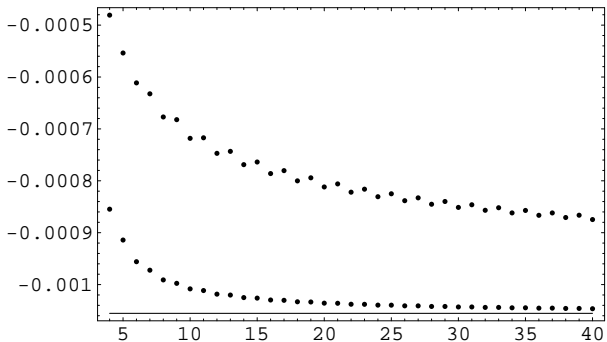}
%\\
%\bpi(0,0)
%\put(11,5){$L$}
%\epi
\end{center}
\vspace{-15pt}
\caption{\label{fig1}
Resummed results for $\De\langle\phi^2\rangle/Nu$ as a function of the
number of loops $L$ for $N\ra\infty$.
Upper dots: LDE.
Lower dots: VPT.
Solid line: exact result.}
\end{figure}

We first treat the $N\ra\infty$ limit, where $\eta=0$ and $\om=\om'=1$
(see, {\em e.g.}, Ref.\ \cite{phi4book}).
For large $N$, the only surviving diagrams are 3-1, 4-1, 5-1, 6-1 and
their generalization to higher orders, which are all easy to compute.
The corresponding resummed results start out as
$f_0^{(4)}=-8.54677\times10^{-4}$, $f_0^{(5)}=-9.14163\times10^{-4}$,
$f_0^{(6)}=-9.55894\times10^{-4}$.
We have plotted $f_0^{(4)},\ldots,f_0^{(40)}$ in Fig.\ \ref{fig1}.
They are seen to converge, albeit slowly, to the exact result 
$\De\langle\phi^2\rangle/Nu=-1/96\pi^2=-1.05543\times10^{-3}$
\cite{BaBlZi}.
We will return to the issue of slow convergence for the $N\ra\infty$
limit below.

The LDE used in Refs.\ \cite{BrRa,SCPiRa1,SCPiRa2,KnPiRa} corresponds to
arbitrarily setting $\om'=2$.
The corresponding values start out as $f_0^{(4)}=-4.80756\times10^{-4}$,
$f_0^{(5)}=-5.53705\times10^{-4}$, $f_0^{(6)}=-6.11288\times10^{-4}$.
We have included $f_0^{(4)},\ldots,f_0^{(40)}$ in Fig.\ \ref{fig1}
to demonstrate the at best very slow convergence.
There are attempts to accelerate it \cite{KnPiRa}, but it has been argued
in Ref.\ \cite{HaKl} that LDE is inapplicable to field theory whenever
there is an irrational exponent of approach to scaling.

Now we turn to the case $N=2$.
We employ two approaches when determining the $r\ra0$ limit of
$\De\langle\phi^2\rangle/Nu$.
One of them uses the known values $\om=0.79\pm0.01$, $\eta=0.037\pm0.003$
(see, {\em e.g.}, Ref.\ \cite{phi4book}) and thus $\om'=0.805\pm0.011$ for
$N=2$, and the other self-consistently determines $\om'$ at each order.
With $\om'=0.805$ (its uncertainty $\pm0.011$ causes a 1\% uncertainty of
the fixed-$\om'$ six-loop estimate for $c_1$) we obtain
$f_0^{(4)}=-4.297\times10^{-4}$, $f_0^{(5)}=-4.814\times10^{-4}$,
$f_0^{(6)}=-5.103\times10^{-4}$, whose translation to $c_1$ via
(\ref{c1}) is plotted in Fig.\ \ref{fig2}.
\begin{figure}[ht]
\begin{center}
% becL6N2plot.eps produced by running ~/bec/math/resum under Mathematica
%\bpi(10,0)
%\put(0,95){$\De\langle\phi^2\rangle/Nu$}
%\epi
\includegraphics[width=8cm,angle=0]{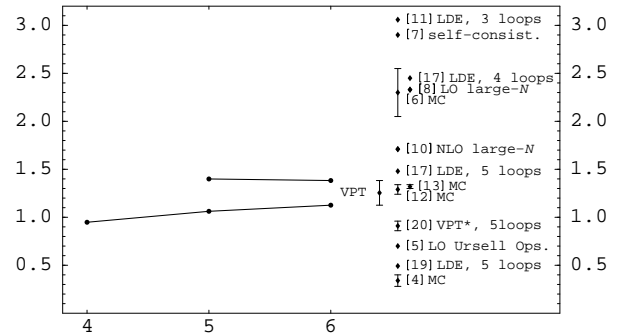}
%\\
%\bpi(0,0)
%\put(11,5){$L$}
%\epi
\end{center}
\vspace{-15pt}
\caption{\label{fig2}
$c_1$ as a function of the number of loops $L$ for $N=2$.
Upper dots: self-consistent $\om'$.
Lower dots: $\om'=0.805$.
The label VPT' indicates the inclusion of a non-zero one-loop
term in Ref.\ \cite{Kl1}, which is absent in the present treatment,
labeled by VPT.
Also included are most results from other sources.
}
\end{figure}

The expected large-$L$ behavior in VPT has the form
$f_0+A\exp[-B(L-3)^{1-\om'}]$ \cite{Kl3,pibook,phi4book}.
However, even and odd loop orders in VPT are, in our case, obtained from
extrema and turning points, respectively, which causes them to behave
slightly differently.
This prohibits a trustworthy extrapolation to $L\ra\infty$ at
the current loop order.
Note that in the large-$N$ limit, where $\om'=1$, the expected large-$L$
behavior turns into $f_0+A'/(L-3)^{B'}$ and an analysis of the data shown
in Fig.\ \ref{fig1}, but extended through $f_0^{(150)}$, yields
$A'\approx3.4\times10^{-4}$, $B'\approx1.0$.

For determining $\om'$ self-consistently \cite{Kl3,phi4book}, we apply VPT
to the quantity $d\ln(\De\langle\phi^2\rangle/Nu)/d\ln u_r$ and exploit
the fact that it vanishes as $u_r\ra\infty$ and that its large-$u_r$
expansion has the same  $\om'$-exponent as $\De\langle\phi^2\rangle/Nu$.
That is, at each loop order we tune $\om'$ such that the optimization
procedure leads to a vanishing approximation for the large-$u_r$ limit
of $d\ln(\De\langle\phi^2\rangle/Nu)/d\ln u_r$.
This $\om'$ is then used at the same loop order in (\ref{f0}).
Our results are $f_0^{(5)}=-6.343\times10^{-4}$ and
$f_0^{(6)}=-6.268\times10^{-4}$, with corresponding $\om'$-values
$\om'^{(5)}=0.6212$ and $\om'^{(6)}=0.6381$.
Via (\ref{c1}) they are converted to the estimates for $c_1$ plotted in
Fig.\ \ref{fig2}.

Taking the average and the difference of our two estimates for
$f_0^{(6)}$ as the mean value and the total error bar of our final result,
respectively, we arrive at
$\De\langle\phi^2\rangle/Nu=-(5.69\pm0.58)\times10^{-4}$ and
via (\ref{c1}) at $c_1=1.25\pm0.13$.
This is in agreement with the recent MC data $c_1=1.32\pm0.02$ \cite{ArMo}
and $c_1=1.29\pm0.05$ \cite{KaPrSv}
(see Refs.\ \cite{ArMo,HoBaBlLa} for explanations of the low MC value in
Ref.\ \cite{GrCeLa}; see Ref.\ \cite{ArMo} for a critique of the MC
calculation of Ref.\ \cite{HoKr}).
Fig.\ \ref{fig2} shows a comparison with most available results.

In summary, we provide a theoretically well-founded resummed perturbative
result for $c_1$, comparing favorably with recent Monte-Carlo data.

The author is grateful to H.~Kleinert for important suggestions, many
discussions,  and a careful reading of the manuscript; also to E.~Braaten
for many helpful comments on the manuscript and to B.~Nickel and T.~Binoth
for useful communications.

\end{document}